\documentclass[usenatbib]{mn2e}

\title[Compact Radio Sources in M82]
  {A Parsec-Scale Study of the 5/15 GHz Spectral Indices of the Compact Radio Sources in M82}
\author[A.~R.~McDonald et~al.]
  {A.~R.~McDonald,$^1$\thanks{email: amd@jb.man.ac.uk} T.~W.~B.~Muxlow$^1$, K.~A.~Wills$^2$, A.~Pedlar$^{1,3}$, R.~J.~Beswick$^1$\\
$^1$University of Manchester, Jodrell Bank Observatory, Macclesfield, Cheshire SK11 9DL.\\
$^2$Department of Physics, University of Sheffield, Sheffield S3 7RH. \\
$^3$Onsala Space Observatory, 439 92 Onsala, Sweden}
\date{Accepted ?. Received ?}

\pagerange{\pageref{firstpage}--\pageref{lastpage}}
\pubyear{2002}

\begin{document}

\maketitle

\label{firstpage}
\begin{abstract}
Observations of the starburst galaxy, M82, have been made with the VLA in its A-configuration at 15~GHz and MERLIN at 5~GHz enabling a spectral analysis of the compact radio structure on a scale of $< 0.1''$ (1.6~pc). Crucial to these observations was the inclusion of the Pie Town VLBA antenna, which increased the resolution of the VLA observations by a factor of $\sim$2. A number of the weaker sources are shown to have thermal spectra and are identified as H{\sc ii} regions with emission measures $\sim$10$^7$~cm$^{-6}$~pc. Some of the sources appear to be optically thick at 5~GHz implying even higher emission measures of $\sim$10$^8$~cm$^{-6}$~pc. The number of compact radio sources in M82 whose origin has been determined is now 46, of which 30 are supernova related and the remaining 16 are H{\sc ii} regions. An additional 15 sources are noted, but have yet to be identified, meaning that the total number of compact sources in M82 is at least 61. Also, it is shown that the distribution of H{\sc ii} regions is correlated with the large-scale ionised gas distribution, but is different from the distribution of supernova remnants. In addition, the brightest H{\sc ii} region at (B1950) 09$^h$ 51$^m$ 42.21$^s$ +69$^{\circ}$ 54$'$ 59.2$''$ shows a spectral index gradient across its resolved structure which we attribute to the source becoming optically thick towards its centre.
\end{abstract}
\section{Introduction}
The radio emission from star-forming galaxies can be either of a thermal (e.g. free-free) or non-thermal (e.g. synchrotron radiation) origin. The non-thermal radiation originates from relativistic electrons which have been accelerated by supernovae resulting from the deaths of massive stars and the thermal emission is from free-free emission from gas which has been ionised by hot, young stars. Therefore, it is obviously of interest to investigate the relative contributions of these processes to the emission which we observe from nearby star-forming galaxies. A separation of the non-thermal and thermal components of radio emission from the archetypal starburst galaxy, M82, has already been carried out by \citet{Allen99} on angular scales of 2$''$ and they identified several complexes of ionised gas. However, 2$''$ is not sufficient resolution to separate the most compact radio sources from the more diffuse radio structure. In this paper we compare observations of M82 made with the VLA in its A-configuration at 15 GHz, incorporating the VLBA Pie Town antenna, with MERLIN 5 GHz observations. The maps made from these observations have a resolution $\leq$~0.1$''$ which extends the analysis of the most compact radio structure into the thermal regime. Thus, we are now able to comment on the relative contribution of thermal and non-thermal processes to the radio emission on scales of order 1~pc. For consistency with previous publications, we assume a distance of 3.2~Mpc to M82 throughout the paper (Burbidge, Burbidge \& Rubin 1964)\nocite{Burbidge64}.

Spectral information on 26 of the compact radio sources in M82 has already been published \citep{Wills97,Allen98}. These analyses were prompted by 408~MHz MERLIN observations and the radio continuum spectra of the detected sources were mainly consistent with non-thermal synchrotron emission, occasionally with a low-frequency turnover which was attributed to free-free absorption by ionised gas. However, free-free absorption is likely to be responsible for the non-detection of many of the remaining sources at low frequencies. In addition, \citet{Wills97} (hereafter W97) identified two sources which were inconsistent with a free-free absorbed synchrotron spectrum, the AGN candidate, 44.01+59.6\footnote{All positions within M82 are quoted relative to (B1950) 09$^h$ 51$^m$ +69$^{\circ}$ 54$'$} and a possible H{\sc ii} region at 40.62+56.0. \citet{Allen98} (hereafter AK98) came to similar conclusions regarding the majority of the sources, but additionally identified a flat-spectrum ($\alpha \sim$ -0.1, where $S \propto \nu^{\alpha}$) source at 42.21+59.2, which they identified as an H{\sc ii} region. However, over 50 compact radio sources have been identified in the nuclear region of M82, therefore until now there has only been spectral information published on around half of the total number, similar to the situation in another nearby starburst, NGC~2146 \citep{Tarchi00}. Also, since previous investigations concentrated on the brightest sources at the longer radio wavelengths, there would have been a bias away from the identification of thermal sources such as H{\sc ii} regions. This is because their electron temperatures are unlikely to exceed 10$^4$~K and hence even if the ionised gas is optically thick, the brightness temperatures cannot exceed this limit (n.b. 10$^4$~K $\equiv$ 0.48~mJy~beam$^{-1}$ with MERLIN at 5~GHz, 50~mas resolution). Therefore, in order to make robust comparisons between the numbers of compact radio sources of different types in starburst galaxies we must first make more confident identifications of a greater fraction of the sources. Furthermore, with the improved sensitivity of radio observations, it is clear that further populations of compact radio sources will be discovered in more distant galaxies. Hence the need to fully investigate the nature of these sources in nearby galaxies.
\section{The observations and data analysis}
 \label{observations}
\begin{figure*}
\begin{center}
\setlength{\unitlength}{1cm}
\begin{picture}(10,17.9)(0,0)
\put(10,0){\includegraphics{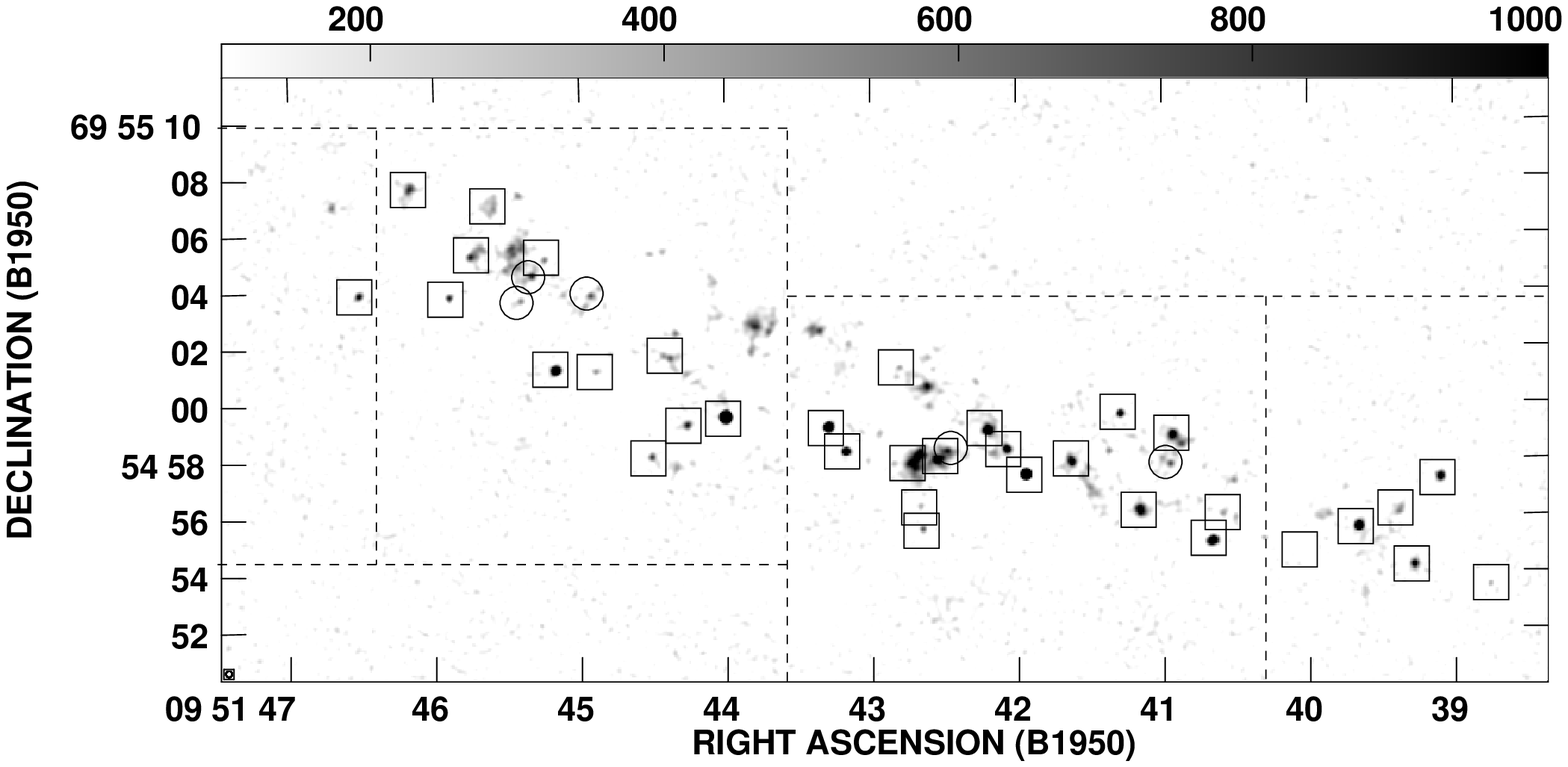}}
\end{picture}
\caption{\label{vlamerlin_compare_fig} VLA A-configuration and Pie Town image of the central regions of M82 at 15~GHz. The grey-scale range is from 0 to 1~mJy~beam$^{-1}$. The dotted lines indicate the fields used to image the MERLIN data and the boxes provide a reference for Figure~\ref{spix_fig} and show the locations within M82 of the spectral index sub-images. Note that this map includes the full spatial frequency range observed by the VLA, hence some of this structure has been `resolved out' in the maps made from the spatial frequency limited data. The boxes indicate which sources are shown in Figure~\ref{spix_fig} due to having significant compact emission at both 5~GHz and 15~GHz and the circles indicate the sources which are shown in Figure~\ref{vlanewfig} since they have significant compact emission at 15~GHz only.}
\end{center}
\end{figure*}

Observations of M82 were performed with the Very Large Array (VLA) of the National Radio Astronomy Observatory, including the Pie Town Very Long Baseline Array (VLBA) telescope, the data from which were correlated in real-time via a fibre-optic link to the VLA site, on the 2nd December 2000. These observations represent the first stage of a programme to image the 15~GHz radio emission from M82 on a wide range of scales and the combination with B- and C-configuation data will be presented in a later paper, along with a discussion of the structures and sizes of the compact sources. However, the A-configuration data alone represent measurements of the radio emission from M82 on scales comparable to those measured using MERLIN at 5~GHz. Therefore, we also present observations made with the MERLIN interferometer in February 1999. The details of both observations are summarised in Table~\ref{obstable}. A number of the parameters listed in Table~\ref{obstable} were chosen to optimise the field of view of the observations since the compact sources in M82 cover a region roughly 50$''$ $\times$ 10$''$ in size. Hence, the VLA correlator was used in spectral line mode, limiting the usable bandwidth to 43.75 MHz, but producing 7 spectral channels with widths of 6.25 MHz. This mode results in smearing at the 5$\%$ level at a distance of 95$''$ from the phase centre of the observations (Taylor, Carilli \& Perley (1999)\nocite{Taylor99}. We deemed this level of smearing to be acceptable for the purposes of our observations and Figure~\ref{vlamerlin_compare_fig} shows the map of the field of view of the observations.
\begin{table*}
\begin{center}
\caption{\label{obstable}Observational setup for the VLA A-configuration (incorporating the VLBA Pie Town antenna) and MERLIN observations of M82.}
\begin{tabular}{|c|c|c|}\hline
Array                                      & VLA (A-configuration) + Pie Town & MERLIN                     \\ \hline
Date                                       & 2nd December 2000                & 5th / 22nd February 1999   \\
Observing Frequency                        & 14.9649 GHz                      & 4.546 / 4.866 / 5.186 GHz  \\
                                           &                                  & (Multi-Frequency Synthesis)\\
Bandwidth                                  & 43.75 MHz                        & 15 MHz                     \\
Number of Frequency Channels               & 7 $\times$ 6.25 MHz              & 15 $\times$ 1 MHz          \\
Integration Time per Visibility            & 3.33 secs                        & 4 secs                     \\
Total Integration Time                     & 12 hours                         & 24 hours                   \\
Longest Baseline (k$\lambda$)              & 3662                             & 3768                       \\
Shortest Baseline (k$\lambda$)             & 20.80                            & 132.1                      \\
Naturally Weighted Sensitivity             & 50                               & 35                         \\
($\mu$Jy beam$^{-1}$)                      &                                  &                            \\
Gaussian {\it u-v} taper FWHM (M$\lambda$) & 4.0                              & 2.2                        \\
{\sc clean} beam size (mas)                & 87$\times$77 @ 58$^{\circ}$      & 83$\times$74 @ 42$^{\circ}$\\
\hline
\end{tabular}
\end{center}
\end{table*}
Table~\ref{obstable} shows that the two sets of observations cover a similar overall range of spatial frequencies, although the VLA has more shorter baselines and hence greater sensitivity to extended structure. In addition, MERLIN has some slightly longer baselines (in terms of wavelengths) than the VLA and Pie Town combination. Therefore, in order to provide a valid comparison, maps were made by limiting both data-sets to a range of 130 k$\lambda < L <$ 3660 k$\lambda$, where $L = (u^2 +v^2)^{1/2}$ and $u$ and $v$ are the co-ordinates of the visibilities in the {\it u-v} plane. The lower limit to this range means that any structures larger than $\sim$1.6$''$ are suppressed. For the purposes of this paper, the word `compact' may describe any source which has significant radio emission on the scales measured by our data, i.e. less than $\sim$1.6$''$ in size. In addition, the density of sampling in the {\it u-v} plane is substantially different between the VLA and MERLIN, since MERLIN is a more sparsely distributed array. Therefore, the relative weighting functions used on the two data-sets needed to be different. The final imaging parameters used in the comparison are also presented in Table~\ref{obstable}.

\begin{figure*}
\begin{center}
\setlength{\unitlength}{1cm}
\begin{picture}(10,11)
\put(-3,0){\includegraphics{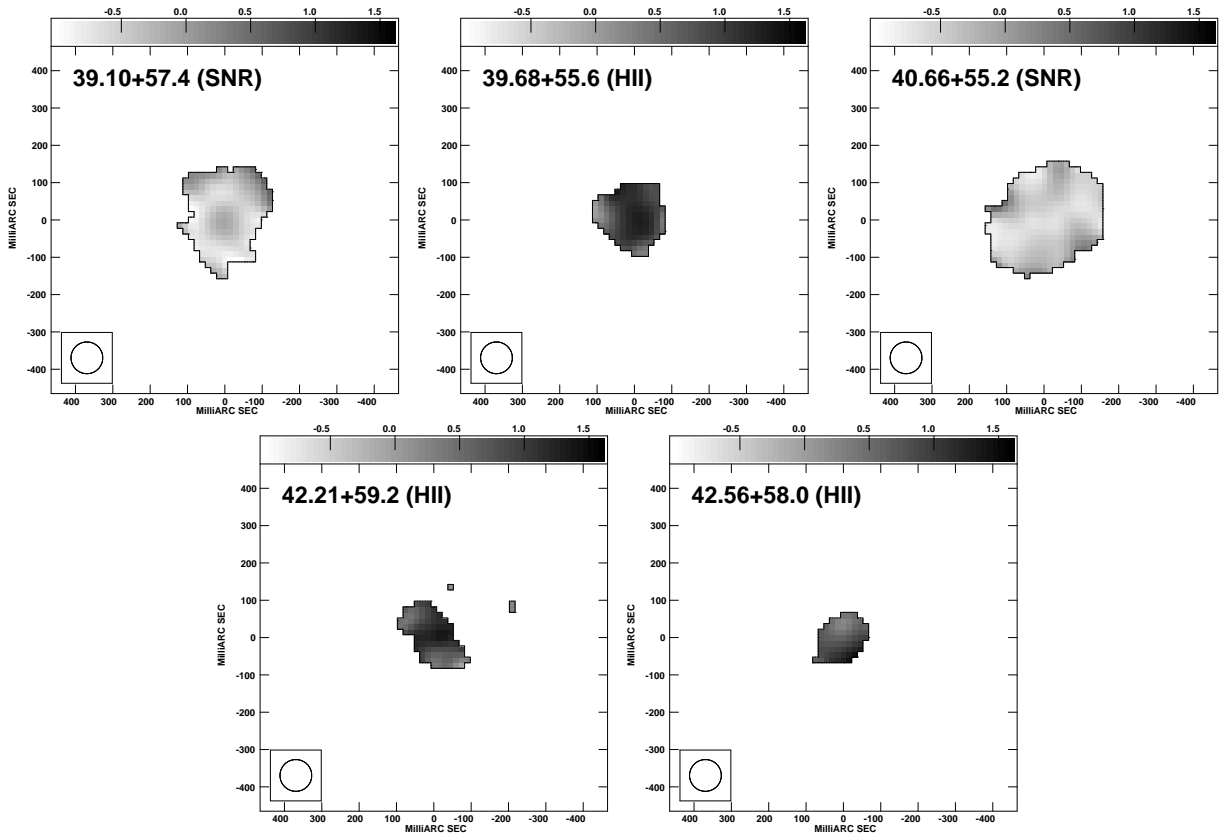}}
\end{picture}
\caption{\label{examplespixfig}Example two-point spectral maps (between 15 GHz and 5 GHz) for 5 of the compact radio sources in M82. The grey-scale range is for -1.0 $< \alpha < $ 1.6.}
\end{center}
\end{figure*}

\begin{figure*}
\begin{center}
\setlength{\unitlength}{1cm}
\begin{picture}(10,11)
\put(-3,0){\includegraphics{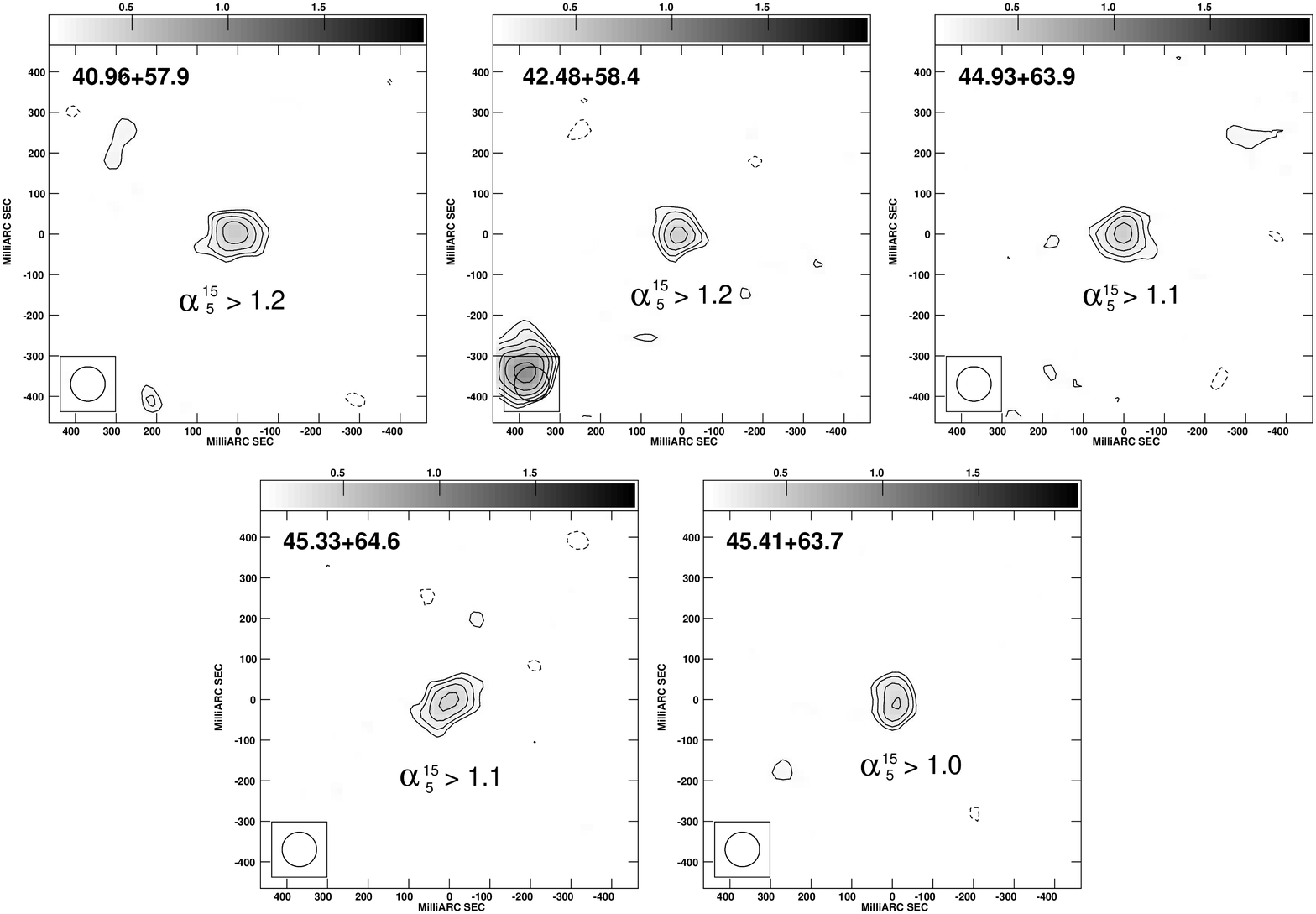}}
\end{picture}
\caption{\label{vlanewfig}VLA A-configuration (incorporating Pie Town) 15 GHz maps of 5 compact sources which remain undetected in the MERLIN 5 GHz images.}
\end{center}
\end{figure*}

The MERLIN and VLA data were reduced separately and self-calibration was used to remove residual complex gain errors after phase-referencing. Therefore, the images from each data-set were registered relative to one another by `fixing' the position of the brightest and most compact radio source, 41.95+57.5, at (B1950) 09$^h$ 51$^m$ 41.95$^s$ +69$^{\circ}$ 54$'$ 57.50$''$ \citep{Pedlar99}. A spectral index map and spectral index error map was then produced for the entire central region of M82 using the {\sc aips} task {\sc comb}. Spectral indices were only calculated for pixels with brightnesses above the 3-$\sigma$ level in each map as stated in Table~\ref{obstable}, using the spectral index convention stated previously. Spectral index error maps were also calculated using the noise levels stated in Table~\ref{obstable}, but are not presented. Sufficient signal was present in the maps at both frequencies such that spectral index maps could be derived for 34 sources. The MERLIN and VLA images of these 34 sources are shown in Figure~\ref{spix_fig} and the spectral index maps for 5 of these sources are shown in Figure~\ref{examplespixfig}. For around 75$\%$ of the 34 sources in Figure~\ref{spix_fig} there exists a good correspondance between the positions of the compact sources (within $\sim$10 mas) at 5~GHz and 15~GHz. In the remaining $\sim$25$\%$ of cases more complex structure is present which does not agree as well between frequencies. Some possible reasons for these discrepancies are discussed in Section~\ref{ID_section}. In addition to the 34 sources detected in both sets of observations, we detect an additional 5 compact sources in the VLA map which have no counterparts in the MERLIN map on the same scale. These sources were chosen by examining the spatial frequency limited VLA 15~GHz map and searching for compact sources with emission above the 5-$\sigma$ level. The VLA maps of these sources are shown in Figure~\ref{vlanewfig} with the corresponding limits on their two-point spectral indices. For reference purposes, Figure~\ref{vlamerlin_compare_fig} contains boxes indicating which sources are illustrated in Figure~\ref{spix_fig} and circles indicating the sources in Figure~\ref{vlanewfig}. It should be noted that Figure~\ref{vlamerlin_compare_fig} shows a significant amount of 15~GHz emission from regions outside of the regions imaged in Figures~\ref{spix_fig} and \ref{vlanewfig}. Most of this emission is suppressed once the limitation on the spatial frequency coverage of the 15~GHz observations is made, although emission is detected at a $\sim$3-$\sigma$ level in these regions.
\begin{figure*}
\begin{center}
\setlength{\unitlength}{1cm}
\begin{picture}(10,21)
\put(-3,0){\includegraphics{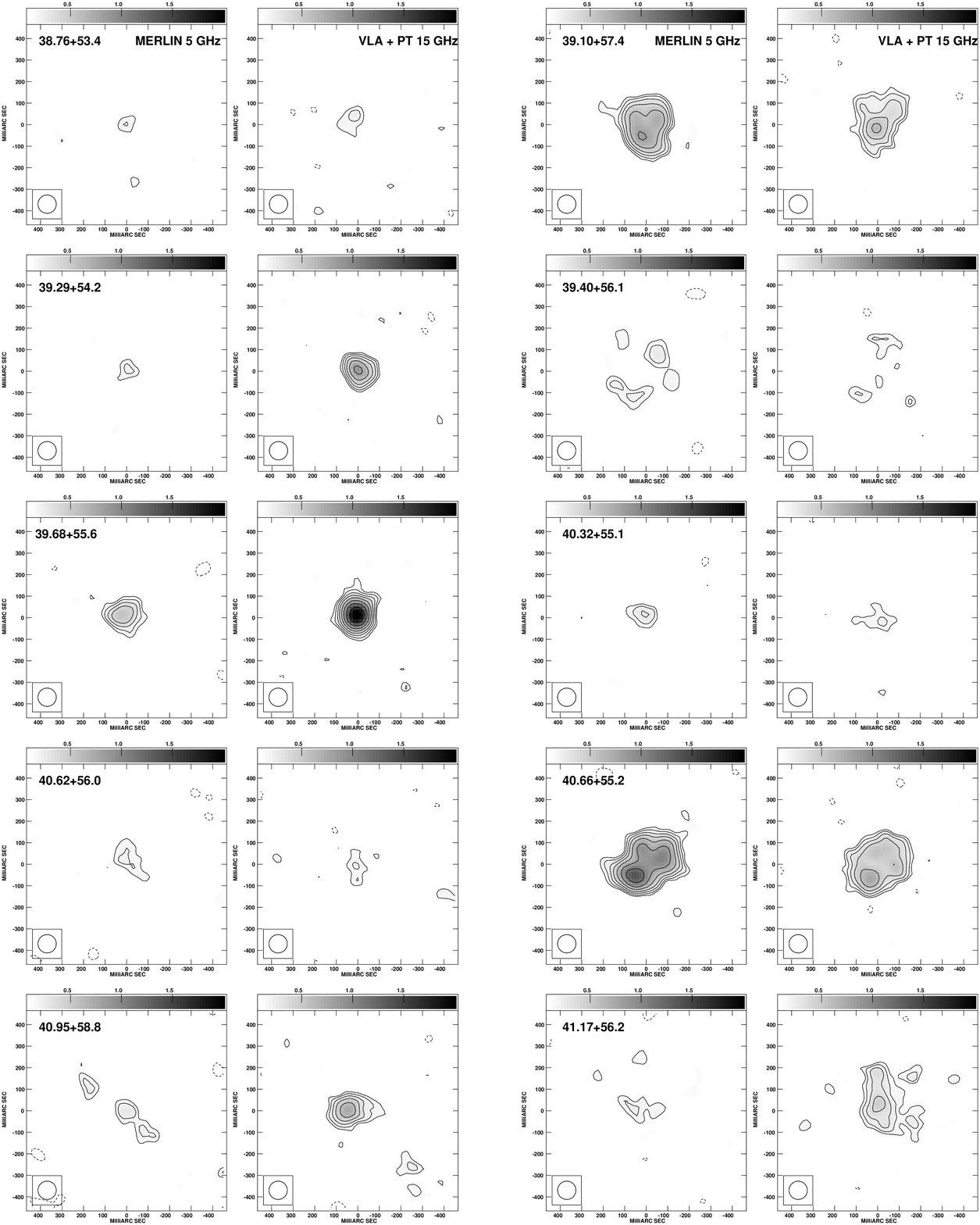}}
\end{picture}
\caption{\label{spix_fig}5 GHz MERLIN and 15 GHz VLA A-configuration (incorporating Pie Town) for 34 compact radio sources in M82 as labelled with boxes in Figure~\ref{vlamerlin_compare_fig}. The contours are at integer multiples of 2$^{\frac{1}{2}}$ above the bottom contour of 140 $\mu$Jy beam$^{-1}$. The grey-scale range is from 0.1 mJy beam$^{-1}$ to 2 mJy beam$^{-1}$.}
\end{center}
\end{figure*}
\begin{figure*}
\begin{center}
\setlength{\unitlength}{1cm}
\begin{picture}(10,21)
\put(0.2,0){\includegraphics{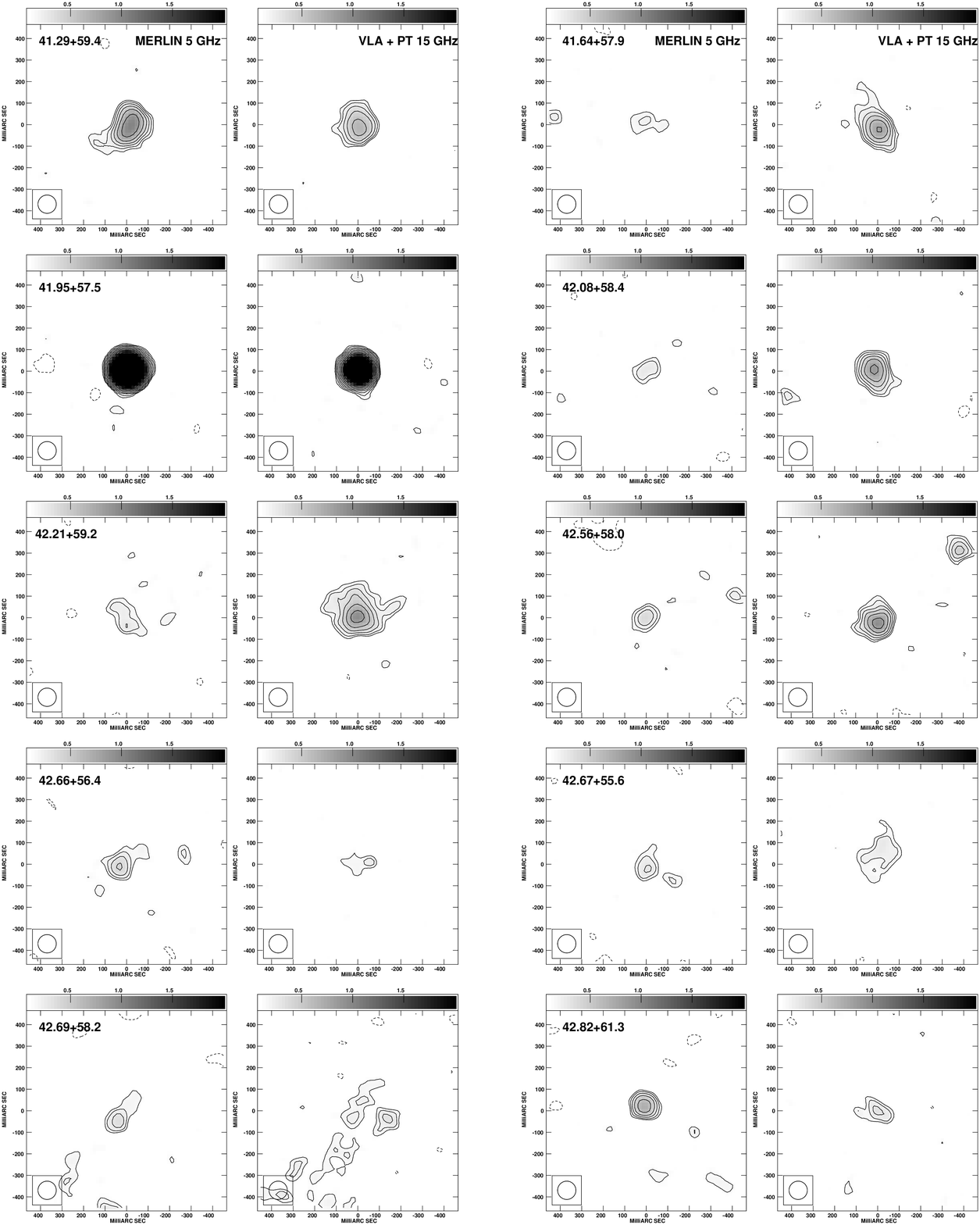}}
\end{picture}
\contcaption{}
\end{center}
\end{figure*}
\begin{figure*}
\begin{center}
\setlength{\unitlength}{1cm}
\begin{picture}(10,21)
\put(0.2,0){\includegraphics{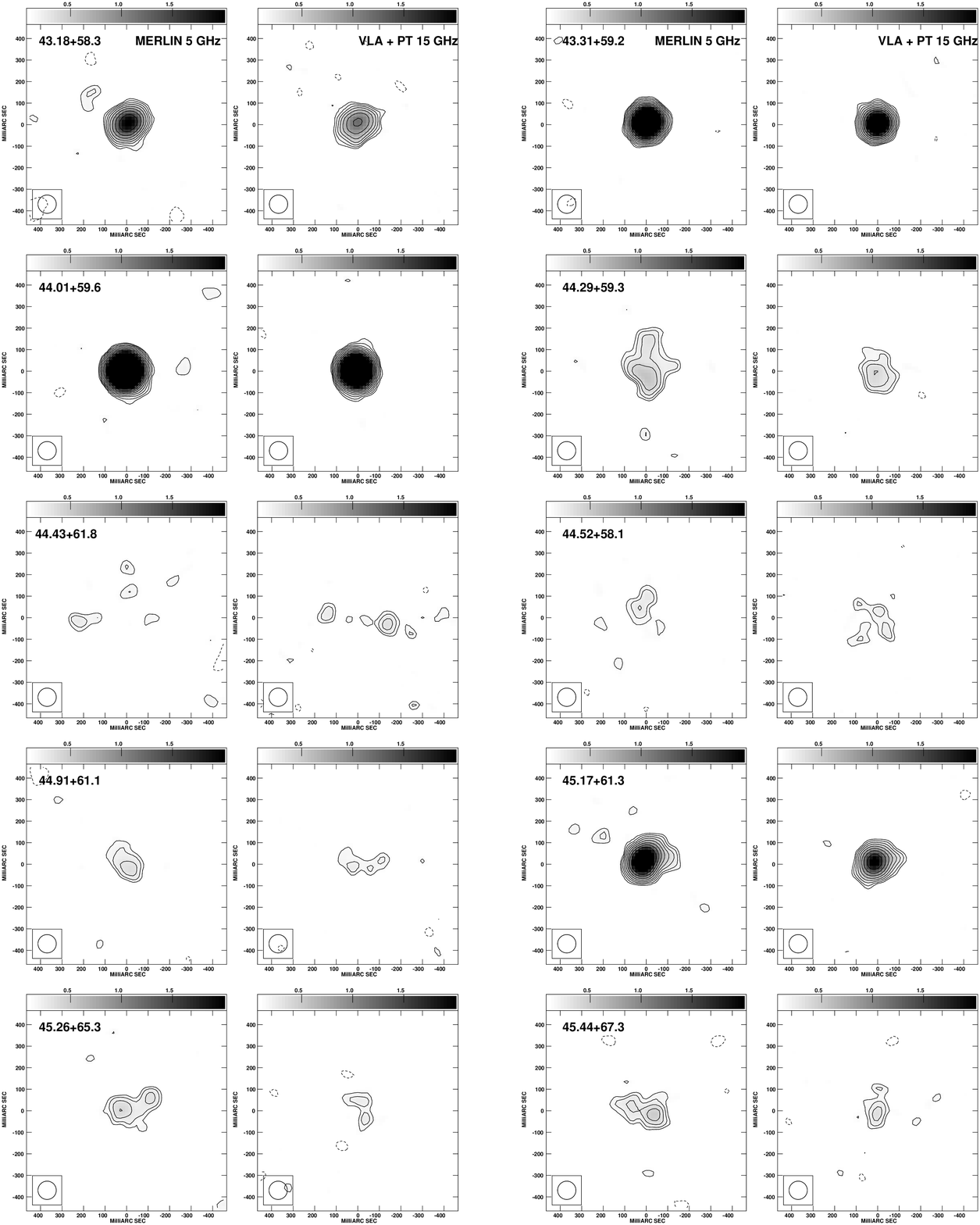}}
\end{picture}
\contcaption{}{}
\end{center}
\end{figure*}
\begin{figure*}
\begin{center}
\setlength{\unitlength}{1cm}
\begin{picture}(10,8)
\put(0.2,0){\includegraphics{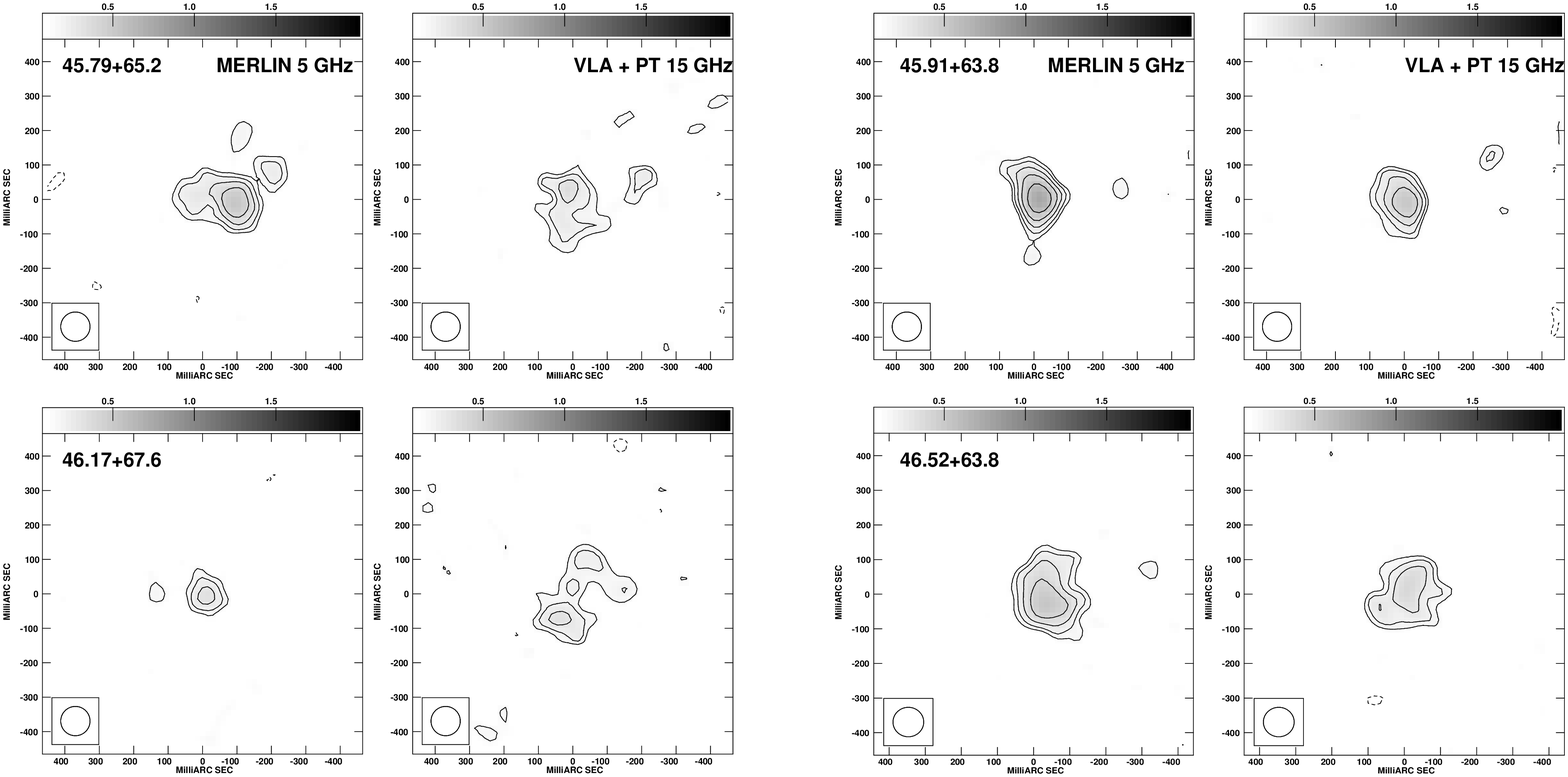}}
\end{picture}
\contcaption{}
\end{center}
\end{figure*}
\section{The nature of the compact sources}
\label{nature_section}
By comparing the maps of the compact radio emission in M82 at 5 GHz and 15 GHz we have detected 34 sources that display compact radio emission at both frequencies as discussed in Section~\ref{observations} and identified an additional 5 compact sources at 15 GHz, but not at 5 GHz. Of these 39 sources, 19 have had spectra published by either W97, AK98 or both. We agree with the original identification of 18 of these 19 sources as being of supernova origin and the remaining source as being an H{\sc ii} region, although the nature of the non-thermal sources at 41.95+59.5 and 44.01+59.6 is still questionable \citep[e.g. ][]{Wills99,Mcdonald01}. Therefore, there were 20 sources which remained to be reliably identified. In order to do this we examined the radio continuum spectrum of each source from 1.4 GHz to 15 GHz.
\subsection{Spectra of the compact sources}
\label{spectra_section}
\begin{figure*}
\begin{center}
\setlength{\unitlength}{1cm}
\begin{picture}(10,18)
\put(-3.5,0){\includegraphics{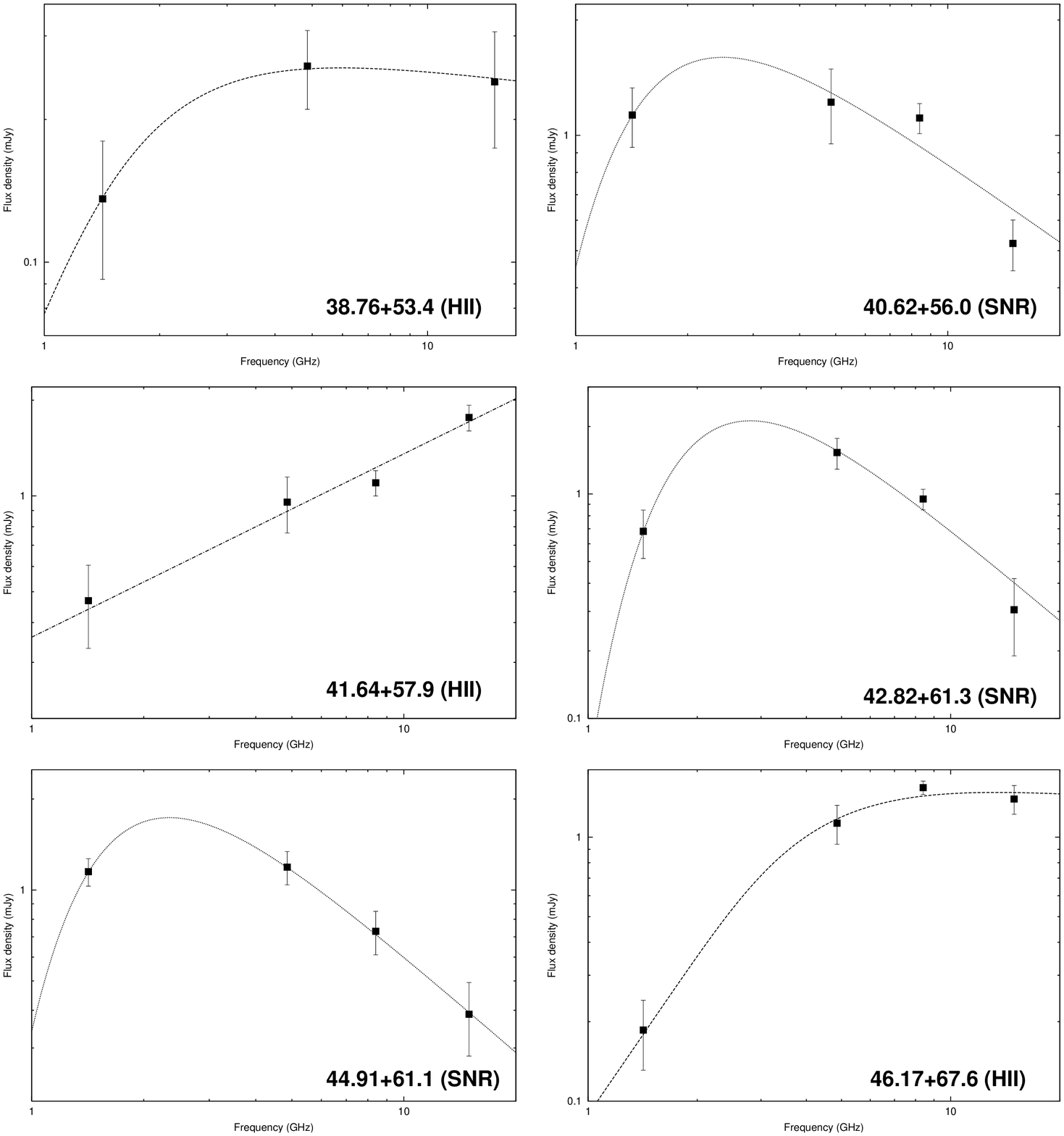}}
\end{picture}
\caption{\label{spectra_fig}Example radio continuum spectra of 6 radio sources in M82 which have not had spectra published previously. The lines indicate the best fits to the integrated flux densities measured from low-resolution ($\sim$0.2$''$) maps.}
\end{center}
\end{figure*}

Since two-point spectral indices only offer a limited amount of information regarding the nature of the source it is obviously of interest to examine the full spectra of as many sources as possible. Therefore, we have used lower resolution ($\sim$0.2$''$ typically) maps at 1.42 GHz (L-band), 4.86 GHz (C-band), 8.4 GHz (X-band) and 15 GHz (U-band) and examined the flux densities at the positions given by the VLA and MERLIN comparison maps.  For the L-band measurements we used the combined VLA and MERLIN map of \citet{Pedlar95} and for the C-band measurements we used an existing VLA A and B-configuration combined image (observed on 5th July 1995 and 31st October 1995 in A and B-configurations respectively). For the X-band measurements we used the data in \citet{Huang94} for the sources available and for the U-band map we used the new VLA data and utilised the full spatial frequency range with an appropriate weighting for a comparison on a scale of $\sim$0.2$''$. Since these low-resolution maps contain a substantial amount of the more diffuse background radio emission in M82, an estimate of the background away from the position of the source was made and an appropriate amount deducted from the measured flux density at the position examined. The errors in the corrected flux densities reflect the uncertainty in the background level which has been subtracted. Table~\ref{fluxtable} shows the measured flux densities of the `new' sources from the high-resolution maps and compares these measurements with the integrated flux densities from the low-resolution maps. Clearly, the flux densities measured from the high-resolution maps are systematically lower than those measured from the low-resolution maps. This discrepancy is partly due to the more diffuse parts of the sources being `resolved out' in the 85 mas maps and to the erroneous inclusion of diffuse background in the measurements from the 200 mas resolution maps.
\begin{table*}
\caption{\label{fluxtable}Integrated flux densities of the radio sources for which new spectral information has been presented in this paper. Measurements made from the low-resolution ($\sim$200 mas) maps and high-resolution ($\sim$85 mas) maps are both presented.}
\begin{tabular}{|c|c|c|c|c|}
\hline
(1)       & (2)             & (3)              & (4)             & (5)\\
Source ID & S$_5$           & S$_{15}$         & S$_5$           & S$_{15}$ \\
          & (85 mas)        & (85 mas)         & (200 mas)       & (200 mas) \\ \hline
38.76+534 & 0.15 $\pm$ 0.03 & 0.27 $\pm$ 0.06  & 0.26 $\pm$ 0.05 & 0.24 $\pm$ 0.07\\
39.29+542 & 0.21 $\pm$ 0.03 & 1.3  $\pm$ 0.1   & 1.7 $\pm$ 0.2   & 1.5 $\pm$ 0.2\\
39.68+556 & 1.1  $\pm$ 0.1  & 3.4  $\pm$ 0.2   & 1.5 $\pm$ 0.2   & 3.2 $\pm$ 0.2\\
40.62+560 & 0.49 $\pm$ 0.09 & 0.22 $\pm$ 0.05  & 1.2 $\pm$ 0.3   & 0.52 $\pm$ 0.08\\
40.95+588 & 0.85 $\pm$ 0.09 & 1.4  $\pm$ 0.1   & 1.4 $\pm$ 0.4   & 2.3 $\pm$ 0.3\\
40.96+579 & -               & 0.62 $\pm$ 0.07  & 0.86 $\pm$ 0.22 & 0.58 $\pm$ 0.08\\
41.17+562 & 0.72 $\pm$ 0.10 & 1.9  $\pm$ 0.2   & 2.0 $\pm$ 0.3   & 2.8 $\pm$ 0.2\\
41.64+573 & 0.36 $\pm$ 0.06 & 1.6  $\pm$ 0.1   & 0.96 $\pm$ 0.19 & 1.8 $\pm$ 0.2\\
42.08+584 & 0.35 $\pm$ 0.06 & 1.6  $\pm$ 0.1   & 1.2 $\pm$ 0.4   & 1.7 $\pm$ 0.2\\
42.48+584 & -               & 0.43 $\pm$ 0.08  & 1.8 $\pm$ 0.2   & 1.0 $\pm$ 0.2\\
42.56+580 & 0.56 $\pm$ 0.07 & 1.5  $\pm$ 0.1   & 2.7 $\pm$ 0.4   & 1.8 $\pm$ 0.2\\
42.66+564 & 0.58 $\pm$ 0.07 & 0.30 $\pm$ 0.06  & 0.62 $\pm$ 0.25 & 0.34 $\pm$ 0.07\\
42.69+582 & 0.54 $\pm$ 0.06 & 1.7  $\pm$ 0.2   & 0.97 $\pm$ 0.43 & 1.8 $\pm$ 0.2\\
42.82+613 & 0.83 $\pm$ 0.08 & 0.41 $\pm$ 0.07  & 1.5 $\pm$ 0.2   & 0.31 $\pm$ 0.12\\
44.43+618 & 0.72 $\pm$ 0.09 & 0.67 $\pm$ 0.10  & 1.1 $\pm$ 0.3   & 0.31 $\pm$ 0.09\\
44.91+611 & 0.71 $\pm$ 0.07 & 0.43 $\pm$ 0.07  & 1.2 $\pm$ 0.2   & 0.39 $\pm$ 0.11\\
44.93+639 & -               & 0.55 $\pm$ 0.08  & 0.26 $\pm$ 0.09 & 0.66 $\pm$ 0.08\\
45.33+646 & -               & 0.59 $\pm$ 0.09  & 0.63 $\pm$ 0.17 & 0.63 $\pm$ 0.11\\
45.41+637 & -               & 0.43 $\pm$ 0.06  & 0.21 $\pm$ 0.07 & 0.48 $\pm$ 0.10\\
46.17+676 & 0.48 $\pm$ 0.06 & 1.0  $\pm$ 0.1   & 1.1 $\pm$ 0.2   & 1.4 $\pm$ 0.2\\
\hline
\end{tabular}
\end{table*}
Figure~\ref{spectra_fig} shows the continuum radio spectra of the 20 sources which have not had spectra previously published by either W97 or AK98. The spectra were fitted by three models for the radio emission and 6 examples are shown in Figure~\ref{spectra_fig}. The models used can be described by the equations
\begin{equation}
S(\nu) = S_0\nu^{\alpha},
\end{equation}
\begin{equation}
S(\nu) = S_0\nu^{\alpha}e^{-\tau(\nu)},
\end{equation}
and
\begin{equation}
S(\nu) = S_0\nu^2(1-e^{-\tau(\nu)}),
\end{equation}
where,
\begin{equation}
\tau(\nu) = (\nu/\nu_{\tau=1})^{-2.1}.
\end{equation}
These equations represent a simple power-law spectrum, a power-law spectrum with free-free absorption by a screen of foreground ionised gas and a self-absorbed Bremsstrahlung spectrum respectively. From this point they will be referred to as models 1, 2 and 3 respectively. In these expressions $\alpha$ is the optically thin spectral index and $\nu_{\tau=1}$ is the frequency at which the free-free optical depth is unity. Of the 20 sources which were modelled, 7 were found to be best fitted by model 1 with an `inverted' spectral index ($\alpha >$ 0), 1 source was best fitted by model 1 with a non-thermal spectral index ($\alpha <$ 0), 4 sources were best fitted by model 2 and 8 sources by model 3. These results are used in the following section to make the final source identifications. Note that we interpret the spectra of the sources which have inverted spectra with a spectral index $<$ 2 (as expected for a simple, optically thick H{\sc ii} region), as being due to an electron density gradient of the type discussed by \citet{Olnon75} and \citet{Panagia75}, since ionised gas with different optical depths which are confused will result in an inverted spectral index (for the entire optically thick source) which is a function of the power-law index of the electron density gradient.
\subsection{Radio source identification}
\label{ID_section}
In order to determine which mechanism is responsible for the radio emission from the compact sources in M82, the morphology of the sources as resolved in the maps shown in Figure~\ref{spix_fig} and the spectral information in Figure~\ref{spectra_fig} were considered. Therefore, in addition to the 26 which have been identified in previous publications, identifications have been made for a further 20 sources. Table~\ref{IDtable} lists all of the 46 sources which have had spectra analysed by W97, AK98 and in this paper as follows.
\begin{enumerate}
\item Source identifier.
\item Position of image origin.
\item Optically thin spectral index deduced by \citet{Wills97} (W97).
\item Optically thin spectral index deduced by \citet{Allen98} - screen absorption model / no absorption (AK98 (a)).
\item Optically thin spectral index deduced by \citet{Allen98} - mixed absorption model (AK98 (b)).
\item Two-point spectral index (this paper).
\item Final identification of source as either supernova-related (SNR), H{\sc ii} region or AGN candidate.
\end{enumerate}
\begin{table*}
\begin{center}
\caption{\label{IDtable}Spectral indices and identifications of the radio sources which have been deduced by \citet{Wills97} (W97), \citet{Allen98} (AK98) and in this paper (M02). Column (iv) contains the spectral indices deduced by AK98 using a screen absorption or no absorption model and column (v) contains the spectral indices deduced by the same authors using a mixed absorption model.}
\begin{tabular}{|c|c|c|c|c|c|c|}
\hline
(i)&(ii)&(iii)&(iv)&(v)&(vi)&(vii)\\
Source ID   & Image Origin & $\alpha$ & $\alpha$ & $\alpha$ & $\alpha_5^{15}$ & ID \\
            & (relative to (B1950) 09$^h$ 51$^m$ +69$^{\circ}$ 54$'$)              & W97                  & AK98(a)                   & AK98(b)                   & M02                   &  \\ \hline
 38.76+53.4 & 38.759 +53.52 &  -                   &  -                        &  -                        &   0.54 $\pm$ 0.27     &  H{\sc ii} \\
 39.10+57.4 & 39.106 +57.33 &  -0.6 $\pm$ 0.1      &  -0.38 $\pm$ 0.01         &  -                        &  -0.53 $\pm$ 0.09     &  SNR \\
 39.29+54.2 & 39.289 +54.25 &  -                   &  -                        &  -                        &   1.64 $\pm$ 0.16     &  H{\sc ii} \\
 39.40+56.1 & 39.391 +56.18 &  -0.50 $\pm$ 0.05    &  -0.21 $\pm$ 0.01         &  -                        &  -1.04 $\pm$ 0.22     &  SNR \\
 39.64+53.4 & -             &  -0.20 $\pm$ 0.05    &  -0.71 $\pm$ 0.01         &  -                        &   -                   &  SNR \\
 39.68+55.6 & 39.668 +55.59 &  -                   &  -                        &  -                        &   1.03 $\pm$ 0.09     &  H{\sc ii} \\
 39.77+56.9 & -             &  -0.50 $\pm$ 0.06    &  -0.49 $\pm$ 0.01         &  -                        &   -                   &  SNR  \\
 40.32+55.1 & 40.317 +55.20 &  -0.50 $\pm$ 0.06    &  -0.55 $\pm$ 0.01         &  -                        &  -0.23 $\pm$ 0.21     &  SNR \\
 40.62+56.0 & 40.594 +56.10 &  -                   &   -                       &  -                        &  -0.72 $\pm$ 0.25     &  SNR \\
 40.66+55.2 & 40.676 +55.11 &  -0.70 $\pm$ 0.05    &   -0.52 $\pm$ 0.01        &  -0.55 $\pm$ 0.01         &  -0.54 $\pm$ 0.08     &  SNR  \\
 40.95+58.8 & 40.938 +58.87 &  -                   &   -                       &  -                        &   0.44 $\pm$ 0.13     &  H{\sc ii}   \\
 40.96+57.9 & 40.955 +57.85 &  -                   &   -                       &  -                        &  $>$ 1.2              &  H{\sc ii} \\
 41.17+56.2 & 41.176 +56.16 &  -                   &   -                       &  -                        &   0.87 $\pm$ 0.14     &  H{\sc ii}  \\
 41.29+59.7 & 41.302 +59.64 &  -                   &   -0.54 $\pm$ 0.01        &  -0.56 $\pm$ 0.01         &  -0.47 $\pm$ 0.09     &  SNR \\
 41.64+57.9 & 41.637 +57.94 &  -                   &   -                       &  -                        &   1.32 $\pm$ 0.18     &  H{\sc ii}  \\
 41.95+57.5 & 41.951 +57.49 &  -0.80 $\pm$ 0.05    &   -0.72 $\pm$ 0.01        &  -0.74 $\pm$ 0.01         &   -                   &  SNR?  \\
 42.08+58.4 & 42.079 +58.39 &  -                   &   -                       &  -                        &   1.32 $\pm$ 0.17     &  H{\sc ii} \\
 42.21+59.2 & 42.210 +59.04 &  -                   &   -0.10 $\pm$ 0.01        &  -                        &   1.16 $\pm$ 0.13     &  H{\sc ii}  \\
 42.48+58.4 & 42.481 +58.36 &  -                   &   -                       &  -                        &  $>$ 1.2              &  H{\sc ii} \\
 42.53+61.9 & -             &  -0.80 $\pm$ 0.05    &   -1.84 $\pm$ 0.01        &  -1.99 $\pm$ 0.01         &   -                   &  SNR \\ 
 42.56+58.0 & 42.557 +58.05 &  -                   &   -                       &  -                        &   0.88 $\pm$ 0.14     &  H{\sc ii}   \\
 42.66+56.4 & 42.673 +56.40 &  -                   &   -                       &  -                        &  -0.58 $\pm$ 0.21     &  SNR  \\
 42.67+55.6 & 42.662 +55.54 &  -0.30 $\pm$ 0.05    &   -0.61 $\pm$ 0.01        &  -0.63 $\pm$ 0.01         &   -1.3 $\pm$ 0.2      &  SNR  \\
 42.69+58.2 & 42.694 +58.24 &  -                   &   -                       &  -                        &   1.04 $\pm$ 0.13     &  H{\sc ii}   \\
 42.82+61.3 & 42.810 +61.30 &  -                   &   -                       &  -                        &  -0.63 $\pm$ 0.17     &  SNR \\
 43.18+58.3 & 43.186 +58.35 &  -0.80 $\pm$ 0.05    &   -0.67 $\pm$ 0.01        &  -0.72 $\pm$ 0.01         &  -0.44 $\pm$ 0.08     &  SNR   \\
 43.31+59.2 & 43.305 +59.20 &  -0.60 $\pm$ 0.06    &   -0.64 $\pm$ 0.01        &  -0.85 $\pm$ 0.01         &  -0.65 $\pm$ 0.07     &  SNR  \\
 44.01+59.6 & 44.008 +59.59 &  0.20 $\pm$ 0.05     &  -0.51 $\pm$ 0.01         &  -0.56 $\pm$ 0.01         &  -0.38 $\pm$ 0.06     &  SNR   \\
 44.29+59.3 & 44.276 +59.30 &   -                  &  -0.56 $\pm$ 0.01         &  -                        &  -0.72 $\pm$ 0.12     &  SNR   \\
 44.43+61.8 & 44.419 +61.72 &   -                  &  -                        &  -                        &   0.07 $\pm$ 0.17     &  SNR  \\
 44.52+58.1 & 44.509 +58.21 &  -0.60 $\pm$ 0.05    &  -0.61 $\pm$ 0.01         &  -0.64 $\pm$ 0.01         &  -0.15 $\pm$ 0.17     &  SNR \\
 44.91+61.1 & 44.899 +61.19 &  -                   &  -                        &  -                        &  -0.45 $\pm$ 0.18     &  SNR  \\
 44.93+63.9 & 44.934 +63.91 &  -                   &  -                        &  -                        &  $>$ 1.1              &  H{\sc ii}   \\
 45.17+61.2 & 45.173 +61.25 &  -0.80 $\pm$ 0.05    &  -0.68 $\pm$ 0.01         &  -0.69 $\pm$ 0.01         &  -0.52 $\pm$ 0.07     &  SNR \\
 45.26+65.3 & 45.254 +65.18 &  -                   &  -0.62 $\pm$ 0.01         &  -                        &  -1.02 $\pm$ 0.20     &  SNR   \\
 45.33+64.6 & 45.330 +64.63 &  -                   &  -                        &  -                        &  $>$ 1.1              &  H{\sc ii}  \\
 45.41+63.7 & 45.409 +63.73 &  -                   &  -                        &  -                        &  $>$ 1.0              &  H{\sc ii}  \\
 45.44+67.3 & 45.423 +67.43 &  -0.60 $\pm$ 0.05    &  -0.57 $\pm$ 0.01         &  -                        &  -0.83 $\pm$ 0.17     &  SNR   \\ 
 45.52+64.7 & -             &  -0.20 $\pm$ 0.09    &  -0.15 $\pm$ 0.01         &  -                        &  -                    &  SNR  \\
 45.79+65.2 & 45.758 +65.32 &  -0.60 $\pm$ 0.05    &  -0.23 $\pm$ 0.01         &  -                        &  -0.55 $\pm$ 0.13     &  SNR  \\
 45.91+63.8 & 45.898 +63.88 &  -0.20 $\pm$ 0.05    &  -0.53 $\pm$ 0.01         &  -0.54 $\pm$ 0.01         &  -0.38 $\pm$ 0.11     &  SNR   \\
 46.17+67.6 & 46.175 +67.72 &  -                   &  -                        &  -                        &   0.66 $\pm$ 0.16     &  H{\sc ii}  \\ 
 46.52+63.8 & 46.521 +63.93 &  -0.40 $\pm$ 0.05    &  -0.73 $\pm$ 0.01         &  -1.50 $\pm$ 0.01         &  -0.35 $\pm$ 0.11     &  SNR   \\
 46.56+73.8 & -             &  -0.60 $\pm$ 0.05    &  -0.78 $\pm$ 0.01         &  -1.04 $\pm$ 0.01         &  -                    &  SNR  \\
 46.75+67.0 & -             &  -0.80 $\pm$ 0.05    &  -0.57 $\pm$ 0.01         &  -0.90 $\pm$ 0.01         &  -                    &  SNR  \\
 47.37+68.0 & -             &  -0.60 $\pm$ 0.07    &  -0.57 $\pm$ 0.01         &  -                        &  -                    &  SNR  \\ \hline
\end{tabular}
\end{center}
\end{table*}
As can be seen from Figure~\ref{spix_fig}, some of the regions imaged contain complex structure, therefore some of the identifications listed in Table~\ref{IDtable} are more robust than others. For example, the sources at 39.29+54.2, 39.68+55.6, 41.64+57.9, 42.08+58.4, 42.21+59.2 and 42.56+58.0 as well as the 5 `new' sources shown in Figure~\ref{vlanewfig} all consist of relatively simple structure which is clearly brighter at 15~GHz than 5~GHz. Hence, these sources are confidently identified as being the most optically thick part of regions of thermal emission. The remaining 5 sources which are listed as H{\sc ii} regions in Table~\ref{IDtable} are located in more morphologically complex regions which are dominated by thermal processes, although a mixture of thermal and non-thermal emission is inevitable at some level. For example, at 5~GHz, 46.17+67.7 has a simple single-component structure which becomes more complex at 15~GHz. A mixture of thermal and non-thermal emission may also be responsible for the structures of sources which have been identified as supernova remnants, examples of which are 44.43+61.8 and 44.52+58.2.

\begin{figure}
\begin{center}
\setlength{\unitlength}{1cm}
\begin{picture}(8,6)
\put(0,0){\includegraphics{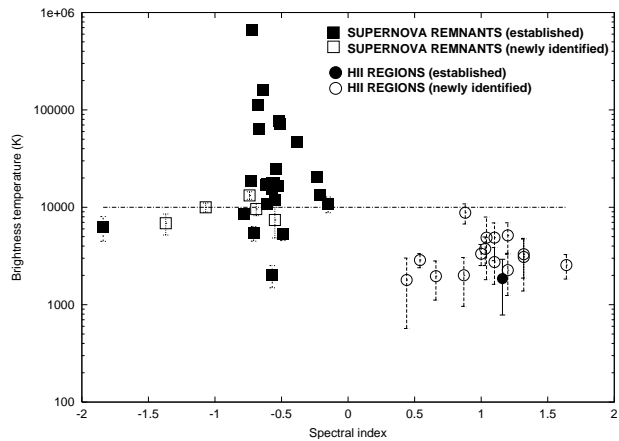}}
\end{picture}
\caption{\label{bias_Tb_fig}Two-point spectral index versus brightness temperature plot for 46 compact radio sources in M82.}
\end{center}
\end{figure}

Of the 26 sources identified previously, only one was deemed to be a possible H{\sc ii} region. This number has now increased to 16 by the addition of 15 sources examined in this paper. In addition, a further 5 supernova remnants have been identified. Therefore, it is suggested that the previous studies were biased towards the identification of supernova remnants, since only the most luminous sources were investigated. Also, the measurements of the source spectra were also biased towards longer wavelengths, away from the regime where thermal sources are brightest - especially if they are optically thick. The observations made with the VLA and Pie Town combination have provided the highest resolution images of the central regions of M82 at 15~GHz and clearly allowed an extension of previous work into the thermal regime.
The peak brightness temperature of a radio source can also be used as a constraint when identifying H{\sc ii} regions. The brightness temperature, $T_b$, is related to the electron temperature, $T_e$ by
\begin{equation}
\label{Tb_equation}
T_b = T_e(1-e^{-\tau}),
\end{equation}
where $\tau$ is the optical depth. Hence in the optically-thick regime ($\tau \gg$ 1) $T_b \approx T_e$. Since ionised gas becomes opaque towards lower frequencies it was decided to measure the brightness temperatures from the lowest frequency map available, which was the L-band (1420 MHz) map of \citet{Pedlar95} and Figure~\ref{bias_Tb_fig} shows the measured brightness temperatures plotted against the source spectral indices. It can be seen that the brightness temperatures of the H{\sc ii} regions all lie below 10$^4$ K, as indicated by the horizontal line. It should be noted that the measured brightness temperature of a thermal source is a lower limit for the electron temperature because as an H{\sc ii} region becomes optically thin the brightness temperature becomes less than the electron temperature. It can also be seen that the brightness temperatures derived for the supernova remnants are mostly in excess of 10$^4$ K, although since the previous studies of the spectra by W97 and AK98 concentrated only on the brightest sources this can be mainly attributed to a selection effect.
\subsection{The radio source distribution in M82}
\begin{figure}
\begin{center}
\setlength{\unitlength}{1cm}
\begin{picture}(8,11)
\put(0.1,0){\includegraphics{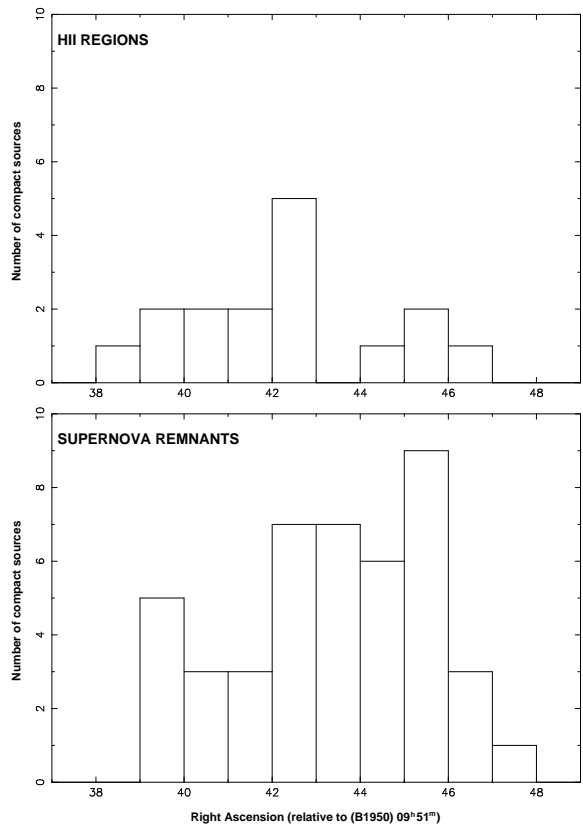}}
\end{picture}
\caption{\label{histogramfig}The one-dimensional distributions of the compact radio sources in M82 projected onto the right ascension axis. The top panel shows the distribution of H{\sc ii} regions and the bottom panel shows the distribution of supernova remnants.}
\end{center}
\end{figure}

\begin{figure}
\begin{center}
\setlength{\unitlength}{1cm}
\begin{picture}(8,6)
\put(-0.2,0){\includegraphics{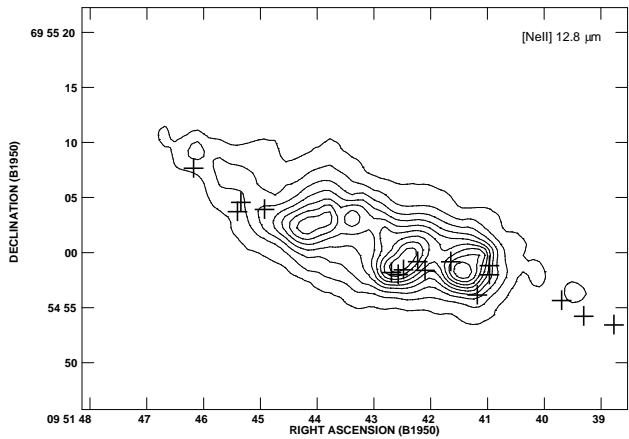}}
\end{picture}
\caption{\label{ionisedgasfig}The ionised gas in M82. The crosses mark the positions of the compact H{\sc ii} regions identified in this paper and are overlayed on the [NeII] line distribution of \citet{Achtermann95}. The contour levels are at integer multiples of 10$\%$ of the peak intensity.}
\end{center}
\end{figure}

Since the number of identified sources in M82 is now 46, a rudimentary statistical analysis of the distributions may be attempted. Comparing the distribution of H{\sc ii} regions with identified supernova remnants, it can be said that the two distributions are different with 80$\%$ confidence. This confidence level was derived using the two-dimensional Kolmogorov-Smirnov test of \citet{Fasano87} and \citet{Peacock83}. Figure~\ref{histogramfig} shows the one-dimensional distributions of the right ascension coordinates of the compact radio sources in M82. The top panel shows the distribution of H{\sc ii} regions and the bottom panel shows the distribution of supernova remnants, including an additional 14 unidentified sources. These sources were noted by \citet{Willsthesis} but were undetected at 408 MHz and were not included in the spectral analysis in W97. However, these sources are almost certainly supernova remnants, since they are generally present on low-frequency ($\sim$1.6 GHz) maps, but not at higher frequencies - implying a non-thermal spectrum. Due to their low brightnesses a spectral analysis has yet to be made for them. In addition, by taking into account the transient source at 41.50+59.7, the total number of compact sources whose origin has been determined is now 61.
The propagation of star-forming regions within M82 has been discussed by \citet{Satyapal97} and \citet{DeGrijs00}. On the basis of a changing CO index across M82 and other indicators of an old stellar population in the nucleus, \citet{Satyapal97} suggested that the starburst was propagating outwards from the centre. However, \citet{DeGrijs00} reported the discovery of `fossil' supernova remnants away from the nucleus ($\sim$ 500 pc away) and suggested that the star-formation was propagating inwards. In order to test these hypotheses with the radio data, the sizes of the H{\sc ii} region and SNR distributions shown in Figure~\ref{histogramfig} was examined, since the H{\sc ii} regions trace the hot, young stars and are more indicative of recent star-formation than the supernova remnants which are the result of star-formation $\sim$10$^7$ years previously. Figure~\ref{ionisedgasfig} shows how the positions of the 16 H{\sc ii} regions correlate with the large scale distribution of ionised gas in M82 as shown by the 12.8 $\mu$m [NeII] map of \citet{Achtermann95}. It can be seen that the [NeII] distribution has a width of $\sim$20$''$ at half-intensity which compares with a width of $\sim$35$''$ for the supernova remnant distribution in Figure~\ref{histogramfig} which may seem to favour the inwardly propagating star-formation hypothesis of \citet{DeGrijs00}. However, small-number statistics in the compact source distribution do not allow a robust distinction to be drawn between the two possibilities.

Again referring to Figure~\ref{ionisedgasfig}, it can be seen that a number of the thermal sources appear to be associated with the two peaks in the [NeII] emission in the west, although a number of the sources seem to be in regions of low [NeII] intensity. However, another two-dimensional Kolmogorov-Smirnov test comparing the compact H{\sc ii} region distribution with the [NeII] line distribution determined that the distributions are similar with a significance level of 90$\%$.
\subsection{\label{nonthermal}Compact radio sources of non-thermal origin}
For the 30 non-thermal sources, it can be seen from Table~\ref{IDtable} that the measured two-point spectral indices are often significantly different from those derived for the models of W97 and AK98. As shown in Figure~\ref{spix_fig}, these sources are often highly resolved and more morphologically complex than the compact H{\sc ii} regions identified previously. It would appear that the two-point spectral indices are more consistent with the models for the brightest, most compact sources and less consistent for the weaker, complex sources. This discrepancy may be partly due to the `resolving out' of larger scale emission. In any case, deciding on what size scale some of the sources `start' is a very subjective matter. 
\subsection{The compact radio sources of thermal origin}
In the previous studies of the spectra of the most compact radio sources in M82, only two sources have been identified as being of possible thermal origin, 40.62+56.0 (W97) and 42.21+59.2 (AK98). It should be noted at this stage that we identify 40.62+56.0 as a supernova remnant due to the extension of the spectrum of this source to higher frequencies than those presented in W97. The spectrum of this source is shown in Figure~\ref{spectra_fig} and if the 15~GHz measurement is excluded then the spectrum agrees with that in W97 and appears to be flat. The higher frequency measurement, however, implies a non-thermal origin for the radio emission from this source. In Section~\ref{ID_section}, 16 sources (including 42.21+59.2) were identified for which positive or flat spectral indices have been derived which are attributable to free-free emission from ionised gas. Table~\ref{HIItable} summarises the inferred properties of these H{\sc ii} regions. For the resolved sources, the sizes were measured by taking a slice through the emission and determining the largest angular size. For the more compact sources, a two-dimensional Gaussian was fitted to the source using the {\sc aips} task {\sc jmfit}. The emission measures were derived from the fits to the spectra of the sources as described in Section~\ref{spectra_section}. For the cases where the sources did not display a turnover in their spectra at the transition between the optically thick and optically thin regimes, a lower limit for the emission measures was taken to be 9 $\times$ 10$^7$ cm$^{-6}$ pc. This limit corresponds to a turnover frequency somewhere above 5~GHz. By calculating the Lyman continuum flux required to ionise the H{\sc ii} regions, given their sizes and emission measures stated in Table~\ref{HIItable}, it is possible to derive an equivalent number of O5 stars, assuming that the Lyman continuum output of an O5 star is $\sim$5.1 $\times$ 10$^{49}$ photons s$^{-1}$ \citep{Spitzer78}. The number of O5 stars required to ionise these regions varies from $\sim$ a few, to over 500. Therefore, these regions may be similar to those found in other galaxies such as NGC~2146 \citep{Tarchi00}, Henize 2-10 \citep{Kobulnicky99} and NGC~5253 (Turner, Beck \& Ho 2000)\nocite{Turner00}, which were deemed to be due to the precursors of `super-star clusters' or `proto-globular clusters'. No optical counterparts were found for the radio sources in these galaxies, however, since these galaxies are not face-on, the optical emission is likely to be obscured by intervening dust and/or the parent molecular cloud.
\begin{table}
\begin{center}
\caption{\label{HIItable}Properties of the H{\sc ii} regions in M82.}
\begin{tabular}{|c|c|c|c|}
\hline
Source     & Size & Emission Measure & Equivalent \\ 
ID         & (pc) & (cm$^{-6}$ pc)   & number of\\ 
           &      & ($\times 10^7 \times (T_e/10^4 K)^{1.35}$) & O5 stars\\ \hline
38.76+53.4 &  1.9 & 1.2              & 3 \\
39.29+54.2 &  1.8 & 2.5              & 6 \\
39.68+55.6 &  5.6 & 21               & 520 \\
40.95+58.8 &  7.2 & 13               & 530 \\
40.96+57.9 &  2.3 & 2.0              & 8 \\
41.17+56.2 &  7.9 & 10               & 490 \\
41.64+57.9 &  5.6 & $>$9.0           & 221 \\
42.08+58.4 &  4.3 & 5.9              & 86 \\
42.21+59.2 &  4.9 & 5.0              & 94 \\
42.48+58.4 &  0.8 & 3.4              & 2 \\
42.56+58.0 &  4.3 & 2.3              & 33 \\
42.69+58.2 &  1.9 & $>$9.0           & 25 \\
44.93+63.9 &  3.7 & $>$9.0           & 97 \\
45.33+64.6 &  2.0 & 1.3              & 4 \\
45.41+63.7 &  1.3 & $>$9.0           & 12 \\
46.17+67.6 &  3.6 & 6.9              & 70 \\ \hline
\end{tabular}
\end{center}
\end{table}

\begin{figure}
\begin{center}
\setlength{\unitlength}{1cm}
\begin{picture}(8,6)
\put(0,0){\includegraphics{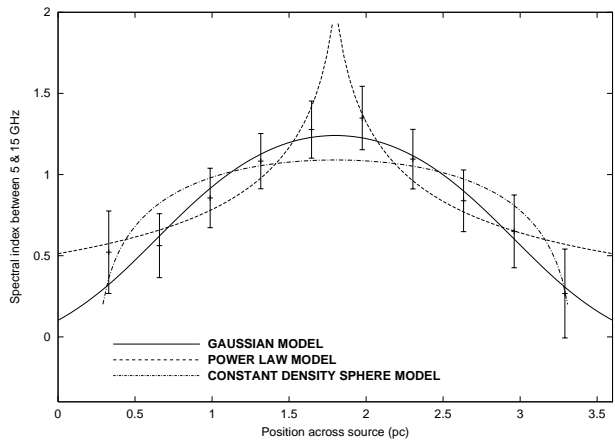}}
\end{picture}
\caption{\label{4221fig}Two-point spectral index (between 5 and 15~GHz) variations across the H{\sc ii} region at 42.21+59.2. The lines indicate the best fits to the profile for 3 different electron density distributions.}
\end{center}
\end{figure}

With the relatively high resolution provided by the VLA and MERLIN (85~mas = 1.3~pc) it is now possible to resolve the compact thermal sources. Variations in the two-point spectral indices can be seen across several of the sources and Figure~\ref{4221fig} shows the spectral index variation across the brightest H{\sc ii} region at 42.21+59.2. Three models for the electron density distribution were fitted to these data - a power-law, Gaussian distribution and constant electron density sphere. As can be seen from Figure~\ref{4221fig}, it is currently not possible to choose between these models. However, the increased spectral index towards the centre of the region may be attributed to increasing optical thickness due to enhanced emission measure.
\section{Conclusions}
\label{discuss}
The VLA in its A-configuration connected with the Pie Town VLBA antenna via a real-time fibre optic link has been used to image M82 at 15 GHz. These observations have provided the highest resolution images yet made at this frequency and allowed a direct comparison with 5 GHz MERLIN images at a similar resolution. The conclusions may be summarised as follows.
\begin{enumerate}
\item In addition to the 26 compact radio sources which have been previously identified in the nucleus of M82, a further 20 sources have been identified. The final identifications are summarised in Table~\ref{final_ID_table} (n.b. the controversial sources at 41.95+57.5 and 44.01+59.5 are included under `supernova remnants'). By comparison with the 12.8 $\mu$m [NeII] map of \citet{Achtermann95}, it was noted that the locations of the compact H{\sc ii} regions broadly coincided with the peaks in the large-scale ionised gas distribution.
\begin{table}
\begin{center}
\caption{\label{final_ID_table}The number of identified compact H{\sc ii} regions and young supernova remnants in M82.}
\begin{tabular}{|c|c|c|c|}\hline
                   & Before                 & This work                     & TOTAL\\
                   & this work              &                               & \\ \hline
Supernova Remnants & 25                     & 5                             & 30   \\
H{\sc ii} Regions  & 1                      & 15                            & 16   \\
\hline
\end{tabular}
\end{center}
\end{table}
\item Of the 15 new H{\sc ii} regions identified, all have steep `inverted' spectra at the highest resolution ($\alpha_5^{15} >$ +0.4). For the simple model of an H{\sc ii} region in which S $\propto \nu^2$ in the optically-thick regime and S $\propto \nu^{-0.1}$ in the optically-thin regime this would imply that these sources are becoming optically thick between 5 and 15 GHz. However, \citet{Olnon75} and \citet{Panagia75} showed that simple models of ionised nebulae with significant electron density gradients could produce inverted spectra, the slope of which depends on the structure of the region. This may be the case for a number of the sources, but further high-resolution imaging at different frequencies is required to confirm or deny this possibility.
\item The low-resolution spectra of the regions in which the compact sources have been imaged show inverted spectra for about half of the sources and flattening spectra at the highest frequencies for the remainder. It is inferred that at the highest resolution only the most compact and optically-thick components of the ionised gas are being detected.
\item Of the 46 identified sources in M82, $\sim$ 35$\%$ are H{\sc ii} regions and the remaining $\sim$ 65$\%$ probably have supernova origins. Therefore, it would appear that M82 is not lacking in compact H{\sc ii} regions in comparison to other galaxies such as NGC~2146 \citep{Tarchi00} as much as was previously thought, although it still appears that M82 is in a more advanced starburst stage than NGC~2146 due to the greater number of supernova remnants.
\item Five additional sources were identified as supernova remnants since they have steep, negative spectral indices at the higher frequencies. At the lower frequencies they are often experiencing a turnover in their spectra by 1.4~GHz requiring emission measures for the foreground ionised gas of $\sim$ 10$^7$ cm$^{-6}$ pc. These supernova remnants are obscured at the lower radio frequencies where previous studies have concentrated and may represent those supernova remnants embedded deeper within M82 and hence behind a greater amount of ionised gas.
\item An additional 14 unidentified sources were added to the analysis and after taking into account the transient source noted by Kronberg, Biermann \& Schwab (1985)\nocite{Kronberg85}, the total number of compact sources now stands at 61. The 14 unidentified sources are most likely the older supernova remnants in the sample and hence are not bright enough to have been included in the essentially flux-limited analyses. However, assuming that they are all SNR, a comparison between the H{\sc ii} region distribution and SNR distribution shows significant differences, although any difference in the overall extent of the distributions is not significant. Hence, the radio data cannot be used to either confirm or deny the propagating star-formation hypotheses of \citet{Satyapal97} and \citet{DeGrijs00}.
\end{enumerate}
\subsection*{Acknowledgements}
MERLIN is a national facility operated by the University of Manchester on behalf of PPARC. The VLA is operated by the National Radio Astronomy Observatory, which is a facility of the National Science Foundation operated under cooperative agreement by Associated Universities, Inc. ARM acknowledges the receipt of a PPARC postgraduate research grant.
\bibliographystyle{mn2e}
\bibliography{papers}

\begin{thebibliography}{}

\bibitem[\protect\citeauthoryear{{Achtermann} \& {Lacy}}{{Achtermann} \&
  {Lacy}}{1995}]{Achtermann95}
{Achtermann} J.~M.,  {Lacy} J.~H.,  1995, \apj, 439, 163

\bibitem[\protect\citeauthoryear{{Allen} \& {Kronberg}}{{Allen} \&
  {Kronberg}}{1998}]{Allen98}
{Allen} M.~L.,  {Kronberg} P.~P.,  1998, \apj, 502, 218

\bibitem[\protect\citeauthoryear{{Allen} \& {Kronberg}}{{Allen} \&
  {Kronberg}}{1999}]{Allen99}
{Allen} M.~L.,  {Kronberg} P.~P.,  1999, in ASP Conf. Ser. 168: New
  Perspectives on the Interstellar Medium p.~205

\bibitem[\protect\citeauthoryear{{Burbidge}, {Burbidge} \& {Rubin}}{{Burbidge}
  et~al.}{1964}]{Burbidge64}
{Burbidge} E.,  {Burbidge} G.,    {Rubin} V.,  1964, \apj, 140, 942

\bibitem[\protect\citeauthoryear{{de Grijs}, {O'Connell}, {Becker}, {Chevalier}
  \& {Gallagher}}{{de Grijs} et~al.}{2000}]{DeGrijs00}
{de Grijs} R.,  {O'Connell} R.~W.,  {Becker} G.~D.,  {Chevalier} R.~A.,
  {Gallagher} J.~S.,  2000, \aj, 119, 681

\bibitem[\protect\citeauthoryear{{Fasano} \& {Franceschini}}{{Fasano} \&
  {Franceschini}}{1987}]{Fasano87}
{Fasano} G.,  {Franceschini} A.,  1987, \mnras, 225, 155

\bibitem[\protect\citeauthoryear{{Huang}, {Thuan}, {Chevalier}, {Condon} \&
  {Yin}}{{Huang} et~al.}{1994}]{Huang94}
{Huang} Z.,  {Thuan} T.,  {Chevalier} R.,  {Condon} J.,    {Yin} Q.,  1994,
  \apj, 424, 114

\bibitem[\protect\citeauthoryear{{Kobulnicky} \& {Johnson}}{{Kobulnicky} \&
  {Johnson}}{1999}]{Kobulnicky99}
{Kobulnicky} H.~A.,  {Johnson} K.~E.,  1999, \apj, 527, 154

\bibitem[\protect\citeauthoryear{{Kronberg}, {Biermann} \& {Schwab}}{{Kronberg}
  et~al.}{1985}]{Kronberg85}
{Kronberg} P.,  {Biermann} P.,    {Schwab} F.,  1985, \apj, 291, 693

\bibitem[\protect\citeauthoryear{{McDonald}, {Muxlow}, {Pedlar}, {Garrett},
  {Wills}, {Garrington}, {Diamond} \& {Wilkinson}}{{McDonald}
  et~al.}{2001}]{Mcdonald01}
{McDonald} A.~R.,  {Muxlow} T. W.~B.,  {Pedlar} A.,  {Garrett} M.~A.,  {Wills}
  K.~A.,  {Garrington} S.~T.,  {Diamond} P.~J.,    {Wilkinson} P.~N.,  2001,
  \mnras, 322, 100

\bibitem[\protect\citeauthoryear{{Olnon}}{{Olnon}}{1975}]{Olnon75}
{Olnon} F.~M.,  1975, \aap, 39, 217

\bibitem[\protect\citeauthoryear{{Panagia} \& {Felli}}{{Panagia} \&
  {Felli}}{1975}]{Panagia75}
{Panagia} N.,  {Felli} M.,  1975, \aap, 39, 1

\bibitem[\protect\citeauthoryear{{Peacock}}{{Peacock}}{1983}]{Peacock83}
{Peacock} J.~A.,  1983, \mnras, 202, 615

\bibitem[\protect\citeauthoryear{{Pedlar} \& {Muxlow}}{{Pedlar} \&
  {Muxlow}}{1995}]{Pedlar95}
{Pedlar} A.,  {Muxlow} T. W.~B.,  1995, \apss, 233, 281

\bibitem[\protect\citeauthoryear{{Pedlar}, {Muxlow}, {Garrett}, {Diamond},
  {Wills}, {Wilkinson} \& {Alef}}{{Pedlar} et~al.}{1999}]{Pedlar99}
{Pedlar} A.,  {Muxlow} T. W.~B.,  {Garrett} M.~A.,  {Diamond} P.,  {Wills}
  K.~A.,  {Wilkinson} P.~N.,    {Alef} W.,  1999, \mnras, 307, 761

\bibitem[\protect\citeauthoryear{{Satyapal}, {Watson}, {Pipher}, {Forrest},
  {Greenhouse}, {Smith}, {Fischer} \& {Woodward}}{{Satyapal}
  et~al.}{1997}]{Satyapal97}
{Satyapal} S.,  {Watson} D.,  {Pipher} J.,  {Forrest} W.,  {Greenhouse} M.,
  {Smith} H.,  {Fischer} J.,    {Woodward} C.,  1997, \apj, 483, 148

\bibitem[\protect\citeauthoryear{{Spitzer}}{{Spitzer}}{1978}]{Spitzer78}
{Spitzer} L.,  1978, {Physical processes in the interstellar medium}.
New York Wiley-Interscience, 1978.~333 p.

\bibitem[\protect\citeauthoryear{{Tarchi}, {Neininger}, {Greve}, {Klein},
  {Garrington}, {Muxlow}, {Pedlar} \& {Glendenning}}{{Tarchi}
  et~al.}{2000}]{Tarchi00}
{Tarchi} A.,  {Neininger} N.,  {Greve} A.,  {Klein} U.,  {Garrington} S.~T.,
  {Muxlow} T. W.~B.,  {Pedlar} A.,    {Glendenning} B.~E.,  2000, \aap, 358, 95

\bibitem[\protect\citeauthoryear{{Taylor}, {Carilli} \& {Perley}}{{Taylor}
  et~al.}{1999}]{Taylor99}
{Taylor} G.~B.,  {Carilli} C.~L.,    {Perley} R.~A.,  eds, 1999, Synthesis
  Imaging in Radio Astronomy II

\bibitem[\protect\citeauthoryear{{Turner}, {Beck} \& {Ho}}{{Turner}
  et~al.}{2000}]{Turner00}
{Turner} J.~L.,  {Beck} S.~C.,    {Ho} P.~T.~P.,  2000, \apjl, 532, L109

\bibitem[\protect\citeauthoryear{Wills}{Wills}{1998}]{Willsthesis}
Wills K.,  1998, PhD thesis, University of Manchester

\bibitem[\protect\citeauthoryear{{Wills}, {Pedlar}, {Muxlow} \&
  {Wilkinson}}{{Wills} et~al.}{1997}]{Wills97}
{Wills} K.,  {Pedlar} A.,  {Muxlow} T.,    {Wilkinson} P.,  1997, \mnras, 291,
  517

\bibitem[\protect\citeauthoryear{{Wills}, {Pedlar}, {Muxlow} \&
  {Stevens}}{{Wills} et~al.}{1999}]{Wills99}
{Wills} K.~A.,  {Pedlar} A.,  {Muxlow} T. W.~B.,    {Stevens} I.~R.,  1999,
  \mnras, 305, 680

\end{thebibliography}
\label{lastpage}
\newpage
\end{document}